# Effects of annealing time on defect-controlled ferromagnetism in $Ga_{1-x}Mn_xAs$

S. J. Potashnik, K. C. Ku, S. H. Chun, J. J. Berry, N. Samarth, and P. Schiffer*

*Department of Physics and Materials Research Institute, The Pennsylvania State University, University Park, PA 16802*

ABSTRACT

We have studied the evolution of the magnetic, electronic, and structural properties of annealed epilayers of $Ga_{1-x}Mn_xAs$ grown by low temperature molecular beam epitaxy. Annealing at the optimal temperature of 250 °C for less than 2 hours significantly enhances the conductivity and ferromagnetism, but continuing the annealing for longer times suppresses both. These data indicate that such annealing induces the defects in $Ga_{1-x}Mn_xAs$ to evolve through at least two different processes, and they point to a complex interplay between the different defects and ferromagnetism in this material.

*schiffer@phys.psu.edu

Heterostructures derived from the ferromagnetic semiconductor $Ga_{1-x}Mn_xAs$ [1,2,3,4] are of great current interest since they enable the incorporation of ferromagnetic elements into important electronic device configurations such as light emitting diodes [5] and resonant tunneling diodes [6]. The spin 5/2 Mn ions act as acceptors, providing holes that mediate a ferromagnetic Mn-Mn coupling. These holes are heavily compensated by defects (e.g. As antisites) created during the low temperature molecular beam epitaxy (MBE) growth of $Ga_{1-x}Mn_xAs$. Such compensation provides a plausible explanation for the observation that the ferromagnetic transition temperature ($T_c$) for $Ga_{1-x}Mn_xAs$ increases with x, reaching a highest reported value of $T_c$ ~110 K for x ~ 0.05 before decreasing again at higher values of x.

The importance of defects to magnetism in $Ga_{1-x}Mn_xAs$ is supported by experimental studies which show that -- even for a fixed Mn composition -- the magnetic properties of $Ga_{1-x}Mn_xAs$ are very sensitive to the detailed MBE growth conditions (growth temperature and beam flux ratios) [7,8] and to post-growth annealing at low temperatures in the range 220 – 310 °C [9]. The latter annealing study showed that $T_c$ can be significantly enhanced over the value for an as-grown sample. Surprisingly, optimal annealing occurs at T ~ 250 °C – close to the growth temperature itself. All these studies signal the complex nature of this material and point to a need for systematic investigations of correlations between its physical properties and growth protocol and post-growth treatments. In this paper, we study the effects of annealing time on the physics of $Ga_{1-x}Mn_xAs$, restricting the annealing to a temperature at which the stoichiometry of the epilayers should be unchanged. We find that ferromagnetism and the conductivity can be enhanced by a relatively short anneal, but is degraded by longer annealing times. These data, combined with changes in the lattice constant and the temperature evolution of the ferromagnetic state, suggest that annealing is associated with at least two different processes among the defects. The results shed light on the role defects play in the ferromagnetism of $Ga_{1-x}Mn_xAs$ and provide a reproducible strategy for maximizing $T_c$.

The $Ga_{1-x}Mn_xAs$ samples in these studies are grown on (001) semi-insulating, epiready GaAs substrates in an EPI 930 MBE system using standard effusion cells and high purity source materials (7N Ga, 7N As and 4N Mn). The substrates are mounted on molybdenum blocks and the temperature during growth (~250$^0$C) is monitored using a radiatively coupled thermocouple situated behind the substrate mounting blocks. Growth is monitored *in situ* with reflection high energy electron diffraction (RHEED) at 12keV. All the $Ga_{1-x}Mn_xAs$ epilayers (with thicknesses in the range 110-140 nm) are grown on a buffer structure that consists of a 120 nm GaAs epilayer grown under standard (high temperature) conditions, followed by a 60 nm low-temperature-grown GaAs epilayer (Ga:As beam equivalent pressure ratio = 1:15). A clear (1x2) reconstruction is observed during growth of the $Ga_{1-x}Mn_xAs$ epilayers upon such templates. After the growth of the $Ga_{1-x}Mn_xAs$ epilayer, all samples are cooled to room temperature in a consistent manner, with insignificant *in situ* annealing at the growth temperature.

The as-grown wafers are cleaved into a number of pieces for systematic post-growth annealing using a Thermolyne 21100 tube furnace in an ultra high purity nitrogen atmosphere flowing at 1.5 standard cubic feet per hour. Magnetization is measured in plane with a commercial superconducting quantum interference device magnetometer in a field of 0.005 T after cooling in a 0.1 T field. Magnetization data taken to T > 310 K show no evidence of MnAs precipitates. X-ray diffraction (XRD) measurements are carried out using a Philips four circle diffractometer with $\lambda=1.54$Å Cu K$\alpha$ X-rays provided by a fixed source tube fitted with a double bounce monochromator. Mn concentrations of the as-grown samples were determined by electron microprobe analysis (EMPA), where the $L$ $\alpha$-lines of Mn, Ga, and As were detected and compared with those of the calibration standards (GaAs and the mineral rhodonite). Use of a low energy electron beam (3 keV) ensured the penetration depth (100 nm) to be shorter than the GaMnAs thickness. Resistivity and Hall measurements are made on Hall bars patterned using conventional photolithography and a chemical wet etch, with electrical contacts made using gold wire leads and indium solder. We emphasize that all the post-growth procedures (removal from In-bonded mounts, Hall bar patterning, electrical contacting) are made in a consistent manner that avoids any additional annealing.

We report measurements on three different series of samples (A, B, and C) grown with slightly different Mn/Ga flux ratios. The lattice constants (from x-ray diffraction), ferromagnetic transition temperatures (taken from the maximum in the slope of M(T)), and Mn concentrations (from EMPA) are shown in Table 1 for the unannealed samples in each series. Although detailed data are shown primarily for series A, the results are qualitatively equivalent for all three series. All anneals in our study were performed at $T_{anneal} \sim 250$ $^0$C with a rapid quench to room temperature at the end of the process.

In figure 1, we show magnetization data as a function of temperature for sample A at several different values of annealing time ($t_{anneal}$). Note that $T_c$ is substantially enhanced by the annealing, from the as-grown value of 65 K up to 101 K for the 0.5 hour anneal. Longer anneals do not, however, continue to enhance the ferromagnetism, and $T_c$ actually drops monotonically with increasing $t_{anneal} > 2$ hours.

Figure 2 shows the resistivity ($\rho_{xx}$) as a function of temperature in sample A for several different values of $t_{anneal}$. We see that $\rho_{xx}(T)$ has a sharp maximum, displaying metallic behavior ($d\rho_{xx}/dT > 0$) below $T_c$, as is typical for ferromagnetic $Ga_{1-x}Mn_xAs$ samples with high $T_c$. We find that $\rho_{xx}$ has a high value for the unannealed sample, decreases with annealing times up to about 2 hours, and then begins to increase again for longer $t_{anneal}$. We also measured the Hall resistance between 10 and 14 tesla at $T = 300$ K, which gives us a qualitative measure of the carrier concentration (the possibility of a contribution from the anomalous Hall effect prevents us from determining the exact value in our field range). The resultant carrier concentration follows the same pattern as the conductivity, first increasing with $t_{anneal}$ and then decreasing for $t_{anneal} > 2$ hours.

Our data reveal two distinct regimes of annealing times as can be clearly seen in figure 3, where we plot several sample properties as a function of $t_{anneal}$. Short anneals of up to 1-2 hours lead to a sharp decrease in the lattice constant and resistivity and corresponding increases in the carrier concentration and the ferromagnetic $T_c$. Moreover, the short anneals change the development of the ferromagnetic ground state as reflected in temperature dependence of the magnetization (Fig. 1). As reported in earlier work, M(T) in as-grown $Ga_{1-x}Mn_xAs$ has a characteristic shape below $T_c$, with a linear rise in M immediately below $T_c$ and then a kink, below which M(T) continues to rise with a smaller slope as $T \to 0$. This behavior has been modeled using a distribution of exchange integrals [10], but we find that annealing for only 10 minutes changes it significantly, moving the kink down to lower temperatures. Anneals of 30 minutes or longer completely transform the qualitative nature of M(T) into a smooth mean-field-like curve typical of other Heisenberg ferromagnets (although both annealed and unannealed samples have the low temperature limiting behavior $\Delta M(T) \propto T^{3/2}$ expected for $Ga_{1-x}Mn_xAs$ [11]). As seen in figure 3, for $t_{anneal}$ longer than about 2 hours, the evolution of the sample properties with annealing changes considerably. These longer annealing times result in a more gradual decrease of the lattice constant and monotonically decreasing values of $T_c$, the low temperature magnetic moment, the carrier concentration, and the conductivity.

These two regimes in annealing time suggest that the samples are evolving through at least two different processes during annealing: the first enhances ferromagnetism and metallicity and the second suppresses them, while both lead to a reduction in the lattice constant. The decrease in lattice constant does not appear to be due to the evaporation of As as had been previously suggested [9], since we obtain the same results from samples which are annealed while capped by another GaMnAs wafer placed on the top surface. This implies that the control of ferromagnetism by annealing is through changes in the defect structure, rather than changes in As stoichiometry.

While we do not at present have detailed microscopic knowledge about the defects in $Ga_{1-x}Mn_xAs$, we can speculate as to the processes possibly responsible for the trends we observe. The primary defects in low-temperature-grown GaAs are As antisites, As interstitials, and Ga vacancies, while in $Ga_{1-x}Mn_xAs$ the Mn ions themselves can also be considered as defects. Annealing presumably leads to a complicated interplay between these different defects, including both clustering and diffusion, as has been demonstrated in detail for GaAs [12]. Furthermore, recent theoretical work on $Ga_{1-x}Mn_xAs$ has shown that the relative positioning of Mn ions and As antisites has a very strong effect on the magnetic properties – leading to possibilities of either ferromagnetic or antiferromagnetic exchanges between the Mn spins [13] which could explain the changes in $T_c$ and the low temperature magnetic moment. One possible explanation of our results [14] is that on short time scales the As antisites diffuse to positions near isolated Mn ions, a situation which would enhance ferromagnetism [13]. On longer timescales, the Mn ions themselves may cluster – leading to the formation of Mn-As complexes with reduced ferromagnetism [13]. Another possibility is that the initial annealing effects are attributable to a reduction in large-scale disorder in the sample, since disorder can also reduce the ferromagnetic coupling [15], while the longer anneals lead to short-range clustering of Mn-As complexes. More detailed modeling of the defects and their dynamics will be needed to ascertain if either of these scenarios can account for the observed behavior.

In summary, we have demonstrated that annealing of $Ga_{1-x}Mn_xAs$ leads to complex changes in its physical properties which evolve with annealing time. These changes are directly attributable to alterations of the structural defects, implying that they play a critical role in the ferromagnetism of this material. A detailed understanding of the defect structure will be important in optimizing the spintronic properties of this material.

We thank Allan MacDonald, Stefano Sanvito, and Nicola Hill for valuable discussions and for sharing unpublished work. N.S., S.H.C., J.J.B., and K.C.K. were supported by grant nos. ONR N00014-99-1-0071 and -0716, DARPA/ONR N00014-99-1-1093. P.S. and S.P. were supported by grant nos. DARPA N00014-00-1-0951, NSF DMR 97-01548

and 01-01318, and ARO DAAD 19-00-1-0125. J.J.B. also acknowledges dissertation fellowship support from the National Science Foundation.

Table 1. Summary of unannealed sample characteristics for our three sample series.

| Sample series | $T_c$(K) | a(Å) | Mn(%) |
|---|---|---|---|
| A | 70 | 5.687 | 8.3±0.8 |
| B | 64 | 5.678 | 5.8±0.6 |
| C | 45 | 5.671 | 3.7±0.4 |

**FIGURE CAPTIONS**

**Figure 1.** The magnetization of sample series A as a function of temperature. Note that while the unannealed sample has a kink in M(T) near 60 K, samples annealed for 30 minutes or more show smooth temperature dependence. The onset of ferromagnetism can be seen at 110 K, equivalent to the highest value reported in the literature.

**Figure 2.** The resistivity of sample series A as a function of temperature. Note the peak in $\rho_{xx}$ near the ferromagnetic $T_c$ and the non-monotonic changes with annealing time.

**Figure 3.** Various sample properties as a function of annealing time: (**a**) Resistivity and carrier concentration of sample series A (estimated from the Hall resistance) at 300 K. (**b**) Lattice constant at 300 K from x-ray diffraction. (**c**) Ferromagnetic transition temperature taken from the maximum in the slope of M(T). In each quantity, there are clearly two regimes of behavior, with ferromagnetism and conductivity optimized at the crossover between the two at annealing times of 1-2 hours.

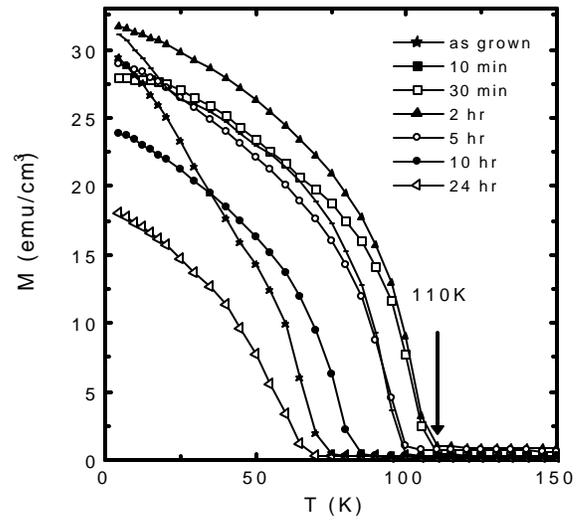

Figure 1. Potashink et al. (APL)

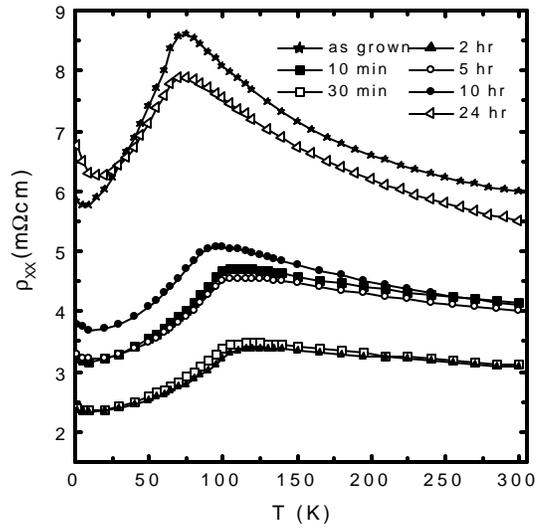

Figure 2. Potashink et al. (APL)

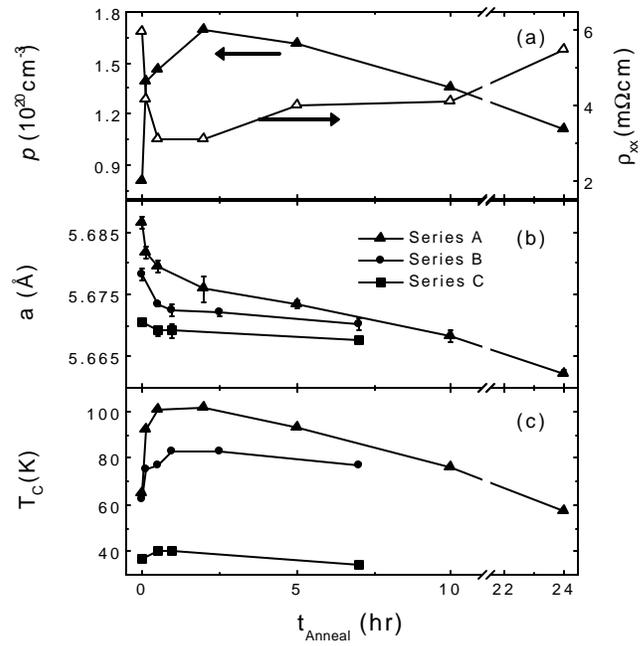

Figure 3. Potashnik et al. (APL)